\documentclass[twocolumn]{aastex63}

\received{}
\revised{}
\accepted{}
\submitjournal{ApJ}

\shorttitle{}
\shortauthors{Deme et al.}
\graphicspath{{./}{figures/}}

\usepackage{amsmath}
\usepackage{appendix}
\usepackage{mathtools}
\usepackage{lineno}

\def\HH#1{{\color{red}\bf #1}}

\begin{document}

\title{A canonical transformation to eliminate resonant perturbations I.}

\correspondingauthor{Barnab{\'a}s Deme}
\email{deme.barnabas@gmail.com}

\author[0000-0003-4016-9778]{Barnab{\'a}s Deme}
\affiliation{Institute of Physics, E\"otv\"os University, P\'azm\'any P. s. 1/A, Budapest, 1117, Hungary}

\author[0000-0002-4865-7517]{Bence Kocsis}
\affiliation{Rudolf Peierls Centre for Theoretical Physics, Clarendon Laboratory, Parks Road, Oxford OX1 3PU, UK}
\affiliation{Institute of Physics, E\"otv\"os University, P\'azm\'any P. s. 1/A, Budapest, 1117, Hungary}

\begin{abstract}

We study dynamical systems which admit action-angle variables at leading order which are subject to nearly resonant perturbations. If the frequencies characterizing the unperturbed system are not in resonance, the long-term dynamical evolution may be integrated by orbit-averaging over the high-frequency angles, thereby evolving the orbit-averaged effect of the perturbations.  
It is well known that such integrators may be constructed via a canonical transformation, which eliminates the high frequency variables from the orbit-averaged quantities. An example of this algorithm in celestial mechanics is the von Zeipel transformation. However if the perturbations are inside or close to a resonance, i.e. the frequencies of the unperturbed system are commensurate, these canonical transformations are subject to divergences. 
We introduce a canonical transformation which eliminates the high frequency phase variables in the Hamiltonian without encountering divergences. This leads to a well-behaved symplectic integrator.  
We demonstrate the algorithm through two examples: a resonantly perturbed harmonic oscillator and the gravitational three-body problem in mean motion resonance.
\end{abstract}


\section{Introduction}\label{intro}

The Kozai--Lidov mechanism \citep{kozai1962,lidov1962} and secular dynamics of triple systems in general, has gained much attention in the past years (see \cite{naoz2016} for a review). This process describes the secular evolution of a hierarchical triple system due to their gravitational interactions, i.e. how a distant tertiary perturbs the dynamics of a binary on timescales much longer than the orbital periods. Its main feature is the presence of a resonant island in phase space, which corresponds to the libration of the binary's argument of periapsis $g_1$ \citep{shevchenko2017}. Hierarchical triples consist of two binaries: the inner binary made up of the tight members and the outer one made up of the tertiary and the barycenter of the inner binary. Kozai--Lidov mechanism (KL hereafter) drives oscillations in the eccentricities of the inner and outer binaries and their mutual inclination, while the semi-major axes remain constant. In the quadrupole approximation ($\propto (a_1/a_2)^2$) the system is integrable (a fact which is referred to as a `happy coincidence' \citealt{lidov1976}). The octupole approximation ($\propto (a_1/a_2)^3$) is chaotic, which drives the eccentricities to almost unity and inclination flipping, a phenomenon called the eccentric KL mechanism \citep{lithwick2011,Katz+2011}. 

The secular equations of motion are derived from the Hamiltonian of the triple system by averaging over the quick angle variables, i.e. the mean anomalies of the inner and outer binaries $l_1$ and $l_2$ \citep{valtonen}. These angles are then cyclic variables in the orbit-averaged effective Hamiltonian, which makes the conjugate momenta constant (i.e. $L_{1,2} \propto \sqrt{a_{1,2}}$, expressed with the semi-major axis), thus the hierarchical three-body problem is stable. Such an elimination of the quick angle variables is carried out by a canonical transformation first applied by von Zeipel \citep{vonzeipel1910}. The von Zeipel transformation is based on a generating function that contains both the original and the new variables, which makes the connection between them implicit and difficult to work with when higher order terms are also taken into account. An alternative derivation utilizes Lie transformations to eliminate the quick angles, which gives explicit connections between the original and the new variables \citep{hori1966}.

However, the orbit-averaged approximation may fail to describe the evolution accurately in many cases
\citep{Liu+2015,Luo+2016,Grishin+2018,Liu_Lai2018,Bhaskar+2021}. In this paper we examine the effects of orbital resonances.

Mean motion resonances (MMRs) play a key role in astrophysics including planetary and stellar dynamics. However, secular evolution in such resonances has been identified to be ``one of the most complicated topics of Celestial Mechanics'' \citep{morbidelli2002}. The main complication is that the generating functions constructed to eliminate the perturbation from the Hamiltonian have divergent denominators in case of MMRs. In other words, the averaged equations of motion obtained by the von Zeipel/Lie transformation break down in the resonant case. The standard way to avoid this problem is to transform the Hamiltonian of the system to new variables, where one of the coordinates is the resonant angle (which changes slowly) and the other is to be averaged over. The new Hamiltonian is analogous to that of a pendulum, where the resonant angle either librates around the exact resonance or it rotates \citep{murray}. In this way, \cite{sansottera2019} reformulate the Laplace--Lagrange theory within mean motion resonances. Another approach was introduced by \cite{wisdom}, a numerical method with which the long-term evolution of perturbed bodies in/near resonances can be efficiently followed. Instead of being eliminated, the quick angle variables are changed in such a way that they sum up as a series of Dirac delta functions in the perturbing Hamiltonian. The dynamics is then driven by either the integrable or the delta-function part of the Hamiltonian: both can be calculated much faster, hence the numerical integration takes $\sim 1000\times$ less CPU time. These ideas were extended to the general $N$-body problem by \cite{Wisdom_Holman1991}.

Keeping the resonant angle in the Hamiltonian results in different equations of motion than the secular ones derived by double-averaging the Hamiltonian. As it is more convenient to solve the same set of equations of motion both in and out of resonances, here we propose a canonical transformation which overcomes the difficulty of small denominators, but contrary to previous studies, we do not use the resonant angle as a canonical variable. Instead, we eliminate the quick angle variables by defining a new set of orbital elements that makes the orbit-averaged dynamical equations valid. As long as the resonant perturbation is small, proportional to some $\epsilon\ll 1$, we demonstrate that the system may be integrated exactly in the transformed variables up to $\epsilon^2$ order, and show that similar subsequent canonical transformations may extend the accuracy of the integration to arbitrary $\epsilon^n$ order. We show that this generates a symplectic integrator for resonant systems.

The paper is structured as follows. In Sec. \ref{sec:general} we describe the canonical transformation with a general Hamiltonian, where we only assume that the perturbation is small and can be decomposed into a convergent Fourier series. In Sec. \ref{sec:toymodel} we apply this transformation to a the case of coupled harmonic oscillators, and in Sec. \ref{sec:triplesystems} to the case of the gravitational three-body problem. We discuss the limitations of the method in Sec. \ref{sec:discussion}.

Throughout the paper, we adopt units where the gravitational constant is $G=1$.

\section{The general secular Hamiltonian in resonance}\label{sec:general}
Let us consider an integrable system, the Hamiltonian, which is expressed with its action variables: $\mathcal{H}_0(\mathbf{J})$. We assume that the system is in resonance or close to it as defined below. Let us also assume that the system is perturbed by a Hamiltonian that can be decomposed into a Fourier series and may be written as
\begin{align}\label{eq:H}
    \mathcal{H}(\mathbf{J},\boldsymbol \theta)
    =&\mathcal{H}_0(\mathbf{J})+\epsilon \sum_\mathbf{m}\mathcal{H}_{1,\mathbf{m}}(\mathbf{J})e^{i\mathbf{m}\cdot\boldsymbol \theta} 
    \nonumber\\=& \mathcal{H}_0(\mathbf{J}) + \epsilon \mathcal{H}_{1,\mathbf{m=0}}(\mathbf{J})+
    \epsilon\sum_{\mathrm{NR},\mathbf{m}}\mathcal{H}_{1,\mathbf{m}}(\mathbf{J})e^{i\mathbf{m}\cdot\boldsymbol \theta}
    \nonumber\\&+\epsilon\sum_{\mathrm{R},\mathbf{m}}\mathcal{H}_{1,\mathbf{m}}(\mathbf{J})e^{i\mathbf{m}\cdot\boldsymbol \theta},
\end{align}
where  $\mathbf{J}$ and $\boldsymbol \theta$ are the action and angle variables (in the absence of perturbations $\mathbf{J}$ are adiabatic invariants), $\mathcal{H}_0$ is the unperturbed Hamiltonian, $\mathrm{R}$ refers to the resonant terms for which $\mathbf{m}$ satisfies $\mathbf{m}\cdot\boldsymbol \omega_0=0$, and $\boldsymbol \omega_0 = \partial \mathcal{H}_0/\partial \mathbf{J}$ are the unperturbed frequencies. $\mathcal{H}_{1,\mathbf{m=0}}(\mathbf{J})$ is the angle-independent part of the perturbing Hamiltonian, which drives the evolution of the system on timescales much longer than the period of $\boldsymbol \theta$ (i.e. on secular timescales $t \gg T_k = 2\pi/\omega_{0,k}$, where $k$ runs through all degrees of freedom). It is obtained by averaging the perturbing part of the Hamiltonian over the angle variables. Physically this procedure amounts to smearing out the orbiting object (e.g. a planet) along its trajectory, which results in a ''mass wire". Technically this is achieved by a canonical transformation for which the transformed momenta contain the effects of the perturbation by definition as we show below. The coupling constant of the perturbation is assumed to satisfy $\epsilon \ll 1$. 

To derive the perturbed secular equations of motion, one has to find a $W$ generating function that eliminates the sums in a way that only the $\mathbf{m}=0$ perturbation remains in the Hamiltonian. In the first order approximation this requirement leads to the so-called homological equation 
\begin{equation}
[\mathcal{H}_0,W]=\epsilon \sum_\mathbf{m} \mathcal{H}_{1,\mathbf{m}}e^{i\mathbf{m}\cdot\boldsymbol \theta},    
\end{equation}
where $[\cdot,\cdot]$ is the Poisson bracket. Its solution is
\begin{equation}\label{generating1}
    W=-\epsilon\sum_\mathbf{m}\frac{\mathcal{H}_{1,\mathbf{m}}e^{i\mathbf{m}\cdot\boldsymbol \theta}}{i\mathbf{m}\cdot\boldsymbol \omega}.
\end{equation}
This famously diverges near MMRs, a phenomenon coined ``the problem of small divisors'' \citep{morbidelli2002}.
Instead of Eq. \eqref{generating1}, we propose the following generating functions in order to eliminate the resonant terms (see \cite{sitaram} for a similar function):
\begin{equation}\label{generating_new1}
    W_k=-\epsilon \frac{\theta_k}{\partial \mathcal{H}_0/\partial J_k}\sum_{\mathrm{R},\mathbf{m}}\mathcal{H}_{1,\mathbf{m}}(\mathbf{J})e^{i\mathbf{m}\cdot\boldsymbol \theta},
\end{equation}
where $k$ can be any of the coordinates. The final results is independent on which $W_{k}$ we use, as we prove below. Eq. \eqref{generating_new1} diverges only when the frequency $\omega_k=\partial \mathcal{H}_0/\partial J_k=0$, but this does not hold at least for the fastest angle variables in celestial mechanics, which is the subject of this study. The generating function of the inverse transformation is 
\begin{equation}\label{generating_new2}
    W'_k=\epsilon \frac{\theta_k'}{\partial \mathcal{H}_0/\partial J_k'}\sum_{\mathrm{R},\mathbf{m}}\mathcal{H}_{1,\mathbf{m}}(\mathbf{J'})e^{i\mathbf{m}\cdot\boldsymbol \theta'},    
\end{equation}
where $\boldsymbol \theta' - \mathbf{J'}$ are the transformed canonical variables.
In what follows, we restrict attention to two degrees of freedom, i. e. $\boldsymbol \theta = (\theta_1, \theta_2)$, $\mathbf{J}=(J_1,J_2)$, to $W_1$ and to the first order $\mathcal{O}(\epsilon)$ approximation. In this case we can substitute the unperturbed quantities into every term which is already multiplied by $\epsilon$, because the difference between $\mathbf{J}$ and $\mathbf{J'}$ is of order $\epsilon$ (see Eqs.~\ref{eq:j1(t)}--\ref{eq:j2(t)}). 

The canonical transformation and its inverse are generated by $W$ and $W'$ for any phase space component $X$ as
\begin{align}
    X' = \exp(L_{W})(X) =& X + \hat{L}_{W}(X) + \frac{1}{2!}\hat{L}_{W}(\hat{L}_{W}(X))
    \nonumber\\&+\dots\\
\label{eq:Jgenerating}
    X = \exp(L_{W'})(X')=& X' + \hat{L}_{W'}(X') + \frac{1}{2!}\hat{L}_{W'}(\hat{L}_{W'}(X'))  
    \nonumber\\&+\dots
\end{align}
where we introduced the Lie operator
\begin{equation}
    \hat{L}_{W'}(\cdot) = [\,\cdot\,, W']\,.
\end{equation}
Substituting Eq.~\eqref{generating_new2} into Eq.~\eqref{eq:Jgenerating} yields 
\begin{align}\label{j1}
    J_1=& \exp(\hat{L}_{W'})(J_1') =  J_1' - \frac{\partial W'}{\partial \theta_1'}+\mathcal{O}(\epsilon^2)\nonumber\\
    =& J_1'- \epsilon \frac{1}{\partial \mathcal{H}_0/\partial J_1'}\sum_{\mathrm{R},\mathbf{m}}\mathcal{H}_{1,\mathbf{m}}e^{i\mathbf{m}\cdot\boldsymbol \theta'} \nonumber\\&- \epsilon \frac{\theta_1'}{\partial \mathcal{H}_0/\partial J_1'}\sum_{\mathrm{R},\mathbf{m}}\mathcal{H}_{1,\mathbf{m}}im_1e^{i\mathbf{m}\cdot\boldsymbol \theta'} + \mathcal{O}(\epsilon^2)\,,\\
\label{j2}
    J_2=&   \exp(\hat{L}_{W'})(J_2') =J_2' - \frac{\partial W'}{\partial \theta_2'}+\mathcal{O}(\epsilon^2)\nonumber\\
    =& J_2' - \epsilon \frac{\theta_1'}{\partial \mathcal{H}_0/\partial J_1'}\sum_{\mathrm{R},\mathbf{m}}\mathcal{H}_{1,\mathbf{m}}im_2e^{i\mathbf{m}\cdot\boldsymbol \theta'}  + \mathcal{O}(\epsilon^2)\,,\\
    \theta_1 =& \exp(\hat{L}_{W'})(\theta_1') =\theta_1' + \frac{\partial W'}{\partial J_1'} +\mathcal{O}(\epsilon^2)\,,\\
    \theta_2 =& \exp(\hat{L}_{W'})(\theta_2') =\theta_1' + \frac{\partial W'}{\partial J_2'} + \mathcal{O}(\epsilon^2)\,.
\end{align}
Note that as long as the system is close to a resonance $\omega_{{\rm res},i}$ for which $\sum_i m_i\omega_{{\rm res},i}=0$ such that there exists $|\epsilon_{\omega i}| \ll 1$ and $m_i$ integers and such that initially 
\begin{align}
\omega_{i,0}&=\frac{\partial \mathcal{H}_0}{\partial J_i} = \omega_{{\rm res},i}(1+\epsilon_{\omega i})\,,
\end{align}
in this case the perturbation terms in Eqs.~\eqref{j1}--\eqref{j2} may be evaluated at resonance as $\partial \mathcal{H}_0/\partial J_1'= \partial \mathcal{H}_0/\partial J_1 + \mathcal{O}(\epsilon)=\omega_{{\rm res},i}+\mathcal{O}(\epsilon,\epsilon_{\omega i})$ since these terms are already multiplied by $\epsilon$ in Eqs.~\eqref{j1}--\eqref{j2}.

The transformed Hamiltonian is equal to the original one expressed with the transformed variables. Substituting Eqs.~\eqref{j1}--\eqref{j2} into the Hamiltonian Eq.~\eqref{eq:H}, and Taylor-expanding with respect to $\epsilon$, and using the fact that for an arbitrary function $F$
\begin{align}\label{eq:F}
    \epsilon F(J') &= \epsilon F(J+\Delta J) 
    = \epsilon F(J) + \epsilon \Delta J \frac{\mathrm{d}F}{\mathrm{d}J}
    + \mathcal{O}(\epsilon^3)
    \nonumber \\ &= \epsilon F(J) + \mathcal{O}(\epsilon^2). 
\end{align}
we get the Hamiltonian in the new variables:
\begin{align}\label{eq:general_hamiltonian}
\begin{array}{clc}
\mathcal{H}' =& 
    \mathcal{H}_0(\exp(\hat{L}_{W'})(\mathbf{J'})) &
    \\&+     \epsilon \sum_{\mathrm{R},\mathbf{m}} \mathcal{H}_{1,\mathbf{m}}e^{i\mathbf{m}\cdot \exp(\hat{L}_{W'})(\mathbf{\boldsymbol\theta'})}&
    \\
    =& \mathcal{H}_0(\mathbf{J'}) & (\mathrm{I}) \\[2ex]
    &-\epsilon\dfrac{\partial \mathcal{H}_0}{\partial J_1'}\dfrac{1}{\partial \mathcal{H}_0/\partial J_1'}\sum_{\mathrm{R},\mathbf{m}}\mathcal{H}_{1,\mathbf{m}}e^{i\mathbf{m}\cdot\boldsymbol \theta'} & (\mathrm{II})\\[2ex] 
    &-\epsilon \dfrac{\partial \mathcal{H}_0}{\partial J_1'}\dfrac{\theta_1'}{\partial \mathcal{H}_0/\partial J_1'}\sum_{\mathrm{R},\mathbf{m}}\mathcal{H}_{1,\mathbf{m}}im_1e^{i\mathbf{m}\cdot\boldsymbol \theta'} & (\mathrm{III}) \\[2ex] 
    &-\epsilon \dfrac{\partial \mathcal{H}_0}{\partial J_2'}\dfrac{\theta_1'}{\partial \mathcal{H}_0/\partial J_1'}\sum_{\mathrm{R},\mathbf{m}}\mathcal{H}_{1,\mathbf{m}}im_2e^{i\mathbf{m}\cdot\boldsymbol \theta'} & (\mathrm{IV}) \\[2ex]
    &+ 
    \epsilon \sum_{\mathrm{R},\mathbf{m}} \mathcal{H}_{1,\mathbf{m}}e^{i\mathbf{m}\cdot\boldsymbol \theta'} + \mathcal{O}(\epsilon^2)
    & (\mathrm{V})
    \\[2ex] 
    \qquad =& \mathcal{H}_0(\mathbf{J'})  + \mathcal{O}(\epsilon^2) & 
\end{array}
\end{align}
Here rows (I-IV) are the transform of $\mathcal{H}_0(J)$ to $\mathcal{O}(\epsilon)$ and row (V) is the perturbation in Eq.~\eqref{eq:H}, which transforms trivially as $\mathbf{J}=\mathbf{J'}$ and $\boldsymbol \theta = \boldsymbol \theta'$ because it is already $\mathcal{O}(\epsilon)$. Rows (II) and (V) trivially cancel each other, while the sum of rows (III) and (IV) vanish if approaching a mean motion resonance:
\begin{align}\label{eq:rows3and4}
    &\frac{-\epsilon\theta_1'}{\partial \mathcal{H}_0/\partial J_1'}\sum_{\mathrm{R},\mathbf{m}}\mathcal{H}_{1,\mathbf{m}}i\left(m_1\frac{\partial \mathcal{H}_0}{\partial J_1'}+m_2\frac{\partial \mathcal{H}_0}{\partial J_2'}\right)e^{i\mathbf{m}\cdot\boldsymbol \theta'} 
    \nonumber\\  
    &=\frac{-\epsilon\theta_1'}{\omega_{1,0}}\sum_{\mathrm{R},\mathbf{m}}\mathcal{H}_{1,\mathbf{m}}i\left(m_1\omega_{1,0}+m_2\omega_{2,0}\right)e^{i\mathbf{m}\cdot\boldsymbol \theta'}
    + \mathcal{O}(\epsilon^2)
    \nonumber\\  
    &= \mathcal{O}(\epsilon \epsilon_{\omega i},\epsilon^2).
\end{align}
What we are left with is finally 
\begin{equation}\label{kamiltonian}
    \mathcal{H}'(\mathbf{J}',\boldsymbol \theta')=\mathcal{H}_0(\mathbf{J}') + \mathcal{O}(\epsilon\epsilon_{\omega i},\epsilon^2),
\end{equation}
which is independent of $\boldsymbol \theta'$ to first order in $\epsilon$ as intended. We note that for the sake of simplicity we omitted the non-resonant sum and the secular term from Eq.~\eqref{eq:H}, because the former only induces small oscillations in the actions, while the latter only results in a small frequency shift. The generating function for the case of both resonant and non-resonant terms is the sum of Eqs. \eqref{generating1} and \eqref{generating_new1}:
\begin{equation}\label{eq:oscplusres}
    W=-\epsilon\sum_{\mathrm{NR,}\mathbf{m}}\frac{\mathcal{H}_{1,\mathbf{m}}e^{i\mathbf{m}\cdot\boldsymbol \theta}}{i\mathbf{m}\cdot\boldsymbol \omega}-\epsilon \frac{\theta_k}{\partial \mathcal{H}_0/\partial J_k}\sum_{\mathrm{R},\mathbf{m}}\mathcal{H}_{1,\mathbf{m}}(\mathbf{J})e^{i\mathbf{m}\cdot\boldsymbol \theta}.
\end{equation}

Such a transformation eliminates the perturbing terms only to $\mathcal{O}(\epsilon)$. Higher order terms may be eliminated in succession to arbitrary order by applying the same procedure. We demonstrate this through an example in Section~\ref{sec:toymodel}.

\subsection{Initial conditions}
Eqs. \eqref{j1} and \eqref{j2} are seemingly asymmetric ($J_1$ has an extra term as a consequence of the arbitrary choice of $\theta_1$ and $J_1$ in Eq. \eqref{generating_new1}), but here we show that the canonical transformation does not have an asymmetry.
$J_1'$ and $J_2'$ are both constant with only
$\mathcal{O}(\epsilon^2)$ corrections
and their values are set by the initial conditions: $(\boldsymbol \theta, \mathbf{J})=(\boldsymbol \theta_0, \mathbf{J_0})$. We assume that the resonance is nearly exact initially, i. e. $\mathbf{m}\cdot\boldsymbol{\omega_0}=\mathcal{O}(\epsilon_{\omega})$. Using Eq.~\eqref{generating_new1} the new $J_1'$ momentum is expressed with the original one as
\begin{align}\label{j1'}
    J_1'=&J_{1,0} + \epsilon \frac{1}{\omega_{1,0}}\sum_{\mathrm{R},\mathbf{m}}\mathcal{H}_{1,\mathbf{m}}e^{i\mathbf{m}\cdot\boldsymbol \theta_0} 
    \nonumber\\ &+ \epsilon \frac{\theta_{1,0}}{\omega_{1,0}}\sum_{\mathrm{R},\mathbf{m}}\mathcal{H}_{1,\mathbf{m}}im_1e^{i\mathbf{m}\cdot\boldsymbol \theta_0} + \mathcal{O}(\epsilon^2),
\end{align}
where $J_{1,0}$ labels the initial value of $J_1$. \footnote{We note that we can replace $\partial \mathcal{H}_0/\partial J_1'$ with $\omega_{1,0}$, because the term including it is already multiplied by $\epsilon$.} Now rewrite Eq. \eqref{j1} as
\begin{align}
    J_1=&J_1'- \epsilon \frac{1}{\omega_{1,0}}\sum_{\mathrm{R},\mathbf{m}}\mathcal{H}_{1,\mathbf{m}}e^{i\mathbf{m}\cdot\boldsymbol \theta_0}
    \nonumber \\
    &- \epsilon \frac{\theta_{1,0}+\omega_{1,0}t}{\omega_{1,0}}\sum_{\mathrm{R},\mathbf{m}}\mathcal{H}_{1,\mathbf{m}}im_1e^{i\mathbf{m}\cdot\boldsymbol \theta_0} + \mathcal{O}(\epsilon^2),
\end{align}
and substitute Eq. \eqref{j1'}, we get
\begin{equation}\label{eq:j1(t)}
    J_1=J_{1,0}(\omega_{1,0})
    - \epsilon t\sum_{\mathrm{R},\mathbf{m}}\mathcal{H}_{1,\mathbf{m}}im_1e^{i\mathbf{m}\cdot\boldsymbol \theta_0} + \mathcal{O}(\epsilon^2).
\end{equation}
Doing the same for Eq. \eqref{j2} yields
\begin{equation}\label{eq:j2(t)}
    J_2=J_{2,0}(\omega_{2,0})-\epsilon t \sum_{\mathrm{R},\mathbf{m}}\mathcal{H}_{1,\mathbf{m}}im_2e^{i\mathbf{m}\cdot\boldsymbol \theta_0} + \mathcal{O}(\epsilon^2).
\end{equation}
These expressions for $J_1$ and $J_2$ are symmetric to the reversal of their index despite the asymmetry caused by the extra term in Eq. \eqref{j1} compared to \eqref{j2}. Note that the neglected terms are expected to be small as long as $t\ll 1/(\epsilon\omega_{i,0})$ for all $i$.

\section{Secular dynamics of coupled harmonic oscillators - a toy model}\label{sec:toymodel}
Here we demonstrate the machinery described in the previous section in a very simple case, where two harmonic oscillators with unit frequency are weakly coupled. The Hamiltonian is
\begin{equation}\label{toymodel_hamiltonian}
    \mathcal{H}=(1+\epsilon_{\omega,1})J_1+
    (1+\epsilon_{\omega,2})J_2 +\epsilon J_1 \sin(\theta_1-\theta_2),
\end{equation}
where $|\epsilon|\ll 1$, $|\epsilon_{\omega 1}|\ll 1$, and $|\epsilon_{\omega 2}|\ll 1$.
The angle variables are the phases of the oscillations and the actions are $J_k=A_k^2/2$, where $A_k$ is the amplitude. The oscillators are weakly coupled as $|\epsilon| \ll 1$. We note that this system is integrable, because it has two first integrals, $\mathcal{H}$ and $J_1+J_2$, whose Poisson bracket vanishes \citep{masoliver2011}. The exact solution is derived in Appendix~\ref{app:analytic_toymodel}. This makes it simple to test the error of our algorithm.

The leading order terms satisfy  $\partial \mathcal{H}_0/\partial J_i=1+\epsilon_{\omega i}$, so $\omega_{i,0}=1+\epsilon_{\omega i}$. This implies that the system is at or close to a 1:1 resonance respectively if $\epsilon_{\omega i}=0$ or $|\epsilon_{\omega i}|\ll 1$ so the standard recipe for eliminating the perturbation diverges, because the perturbing term depends on the difference of the angle variables $\theta_1-\theta_2$ which results in $\mathbf{m}\cdot \boldsymbol \omega = \omega_{1,0}-\omega_{2,0}=\epsilon_{\omega 1}\omega_{1,0} - \epsilon_{\omega 2}\omega_{2,0}$ approaching zero in the denominator of Eq. \eqref{generating1}.

\subsection{First order approximation}
First we eliminate the terms proportional to $\epsilon$. Let us define the generating function using Eqs. \eqref{generating_new1} and \eqref{generating_new2} with $k=1$, which simplify to
\begin{align}
    W&=-\epsilon_0\theta_1 J_1\sin(\theta_1-\theta_2)\,,\\
    W'&=\epsilon_0 \theta_1'J_1' \sin(\theta_1'-\theta_2')\,.
\end{align}
where
\begin{equation}
    \epsilon_0 = \frac{\epsilon}{1+\epsilon_{\omega1}}\,.
\end{equation}
The canonical transformation formulae between the original and the new variables are given by Eqs.~\eqref{j1}--\eqref{j2}, i.e.
\begin{align}\label{eq:toymodel1a}
    J_1=&e^{\hat L_{W'}}J_1' = J_1'  -\epsilon_0  J_1'\theta_1' \cos \theta' - \epsilon_0 J_1' \sin \theta'\nonumber \\ &+\frac{\epsilon_0^2}{2}J_1'\theta_1'^2+\frac{\epsilon_0^2}{2}J_1'\sin^2\theta'+ \mathcal{O}(\epsilon_0^3)
    \\
    J_2=&e^{\hat L_{W'}}J_2' = J_2'+\epsilon_0 J_1'\theta_1'\cos\theta'-\frac{\epsilon_0^2}{2}J_1'\theta_1'^2+\mathcal{O}(\epsilon_0^3),\\
    \theta_1=&e^{\hat L_{W'}}\theta_1' = \theta_1' + \epsilon_0 \theta_1'\sin \theta'+ \frac{\epsilon_0^2}{4} \theta_1'^2\sin(2\theta') 
    \nonumber\\
    &+ \frac{\epsilon_0^2}{2} \theta_1'\sin^2\theta' + \mathcal{O}(\epsilon_0^3),\\
    \theta_2=&\theta_2',
    \label{eq:toymodel1d}
\end{align}
where $\theta'= \theta_1'-\theta_2'$. 

The new Hamiltonian may be obtained by substituting into Eq.~\eqref{toymodel_hamiltonian} or by $\mathcal{H}'=e^{\hat L_W}\mathcal{H}$:
\begin{align}\label{transformed}
    \mathcal{H}'(\boldsymbol{\theta'},\mathbf{J'}) =&
    (1+\epsilon_{\omega 1})J_1' + (1+\epsilon_{\omega 2})J_2'
    -\frac{\epsilon\epsilon_0}{2} J_1'\sin^2 \theta'
        \nonumber \\
    &-\epsilon_0\epsilon_{\omega} J_1'\theta_1'\cos\theta'
     + \mathcal{O}(\epsilon_0^3,\epsilon_0^2\epsilon_{\omega 1})
    \nonumber \\
        =& \left(1+\epsilon_{\omega 1}-\frac{\epsilon_0^2}{4}\right)J_1' + (1+\epsilon_{\omega 2}) J_2'  
        \nonumber \\
    &+\frac{\epsilon_0^2}{4}J_1'\cos 2\theta' 
    - \epsilon_0\epsilon_{\omega} J_1' \theta_1'\cos\theta'
    \nonumber \\
     &+ \mathcal{O}(\epsilon_0^3,\epsilon_0^2\epsilon_{\omega 1})
\end{align}
where we introduced the notation
\begin{equation}
    \epsilon_{\omega} =\epsilon_{\omega 1}-\epsilon_{\omega 2}\,.
\end{equation}
Neglecting the $\mathcal{O}(\epsilon_0^2)$ corrections, the equations of motion of the primed variables are formally the same as the unperturbed/averaged ones:
\begin{align}\label{eq:j1prime}
    \dot{J_1'}&=0 + \mathcal{O}(\epsilon_0^2), \\ 
    \label{eq:j2prime}
    \dot{J_2'}&=0 + \mathcal{O}(\epsilon_0^2), \\
    \label{eq:theta1prime}
    \dot{\theta_1'}&=1 + \epsilon_{\omega 1} +  \mathcal{O}(\epsilon_0^2), \\
    \label{eq:theta2prime}
    \dot{\theta_2'}&=1 + \epsilon_{\omega 2} + \mathcal{O}(\epsilon_0^2).
\end{align}
As the perturbation is second order, $(J_1', J_2')$ are conserved if neglecting  $\mathcal{O}(\epsilon_0^2)$ perturbations. The $\epsilon_0^2$ corrections 
include both a secular ($-\frac14\epsilon_0^2 J_1'$) and a resonant term ($\frac14\epsilon_0^2J_1'\cos 2\theta' + \epsilon_0\epsilon_{\omega} J_1' \theta_1'\cos\theta'$). 
The second-order orbit-averaged evolution corresponds to dropping the periodic term, however this simplification is not necessary as shown in the next subsection. 
The secular term results in a constant secular shift in the frequency of the first oscillator since $\omega_1'(t)= \partial\mathcal{H}'/\partial J_1' = 1+ \epsilon_{\omega 1}-\frac{1}{4}\epsilon_0^2 + \mathcal{O}(\epsilon^3)$. The frequencies of the oscillators may change secularly for more general perturbations.

\subsection{Second order approximation}
Let us now proceed to  eliminate the remaining perturbation terms $\frac14\epsilon_0^2 J_1'\cos 2\theta'-\epsilon_0 \epsilon_{\omega} J_1' \theta_1'\cos\theta'$ from the Hamiltonian \eqref{transformed} to $\epsilon^2$ order. 
For $\mathcal{H}_0 = (1+\epsilon_{\omega 1}-\frac14\epsilon_0^2)J_1' + (1+\epsilon_{\omega 2})J_2'$, the generating function is chosen using Eqs. \eqref{generating_new1} and \eqref{generating_new2}, which simplifies to
\begin{align}
    W''=
    \frac{\epsilon_{00}\epsilon_0}{4} J_1'' \theta_1''\cos 2\theta''  
    -
    \epsilon_{00} \epsilon_{\omega}J_1''\theta_1''\theta''\cos\theta''\,,
\end{align}
where $\theta''= \theta_1''-\theta_2''$ and
\begin{equation}
    \epsilon_{00} = \frac{\epsilon_{0}}{1+\epsilon_{\omega 1} -\frac14\epsilon_0^2} = \frac{\epsilon}{(1+\epsilon_{\omega 1})(1+\epsilon_{\omega 1} -\frac14\epsilon_0 \epsilon)}
\end{equation}
The new canonical variables are
\begin{align}\label{eq:toymodel2a}
    J_1'=&J_1''-\frac{\epsilon_{00}\epsilon}{4}J_1''(\cos2\theta'' - 2\theta_1''\sin2\theta'')\nonumber\\
    &+
    \epsilon_{00} \epsilon_{\omega}
    J_1''[(\theta_1''+\theta'')\cos\theta''
    -
    \theta_1''\theta'' \sin\theta'']   \nonumber\\
    &+\mathcal{O}(\epsilon^4,\epsilon^2\epsilon_{\omega}^2)\nonumber\\
    J_2'=& J_2'' - \frac{\epsilon_{00}\epsilon}{2}\theta_1''J_1''\sin2\theta''
    \nonumber\\&
    +\epsilon_0 \epsilon_{\omega}
    J_1''\theta_1''(\theta''\sin\theta''
    -\cos\theta'')   
    +\mathcal{O}(\epsilon^4,\epsilon^2\epsilon_{\omega}^2)\nonumber\\
    \theta_1'=&\theta_1'' +
    \frac{\epsilon_{00}\epsilon}{4}\theta_1''\cos2\theta''
    -
        \epsilon_{00} \epsilon_{\omega}\theta_1''\theta''\cos\theta''
    \nonumber\\&
    +\mathcal{O}(\epsilon^4,\epsilon^2\epsilon_{\omega}^2)\\
    \theta_2'=&\theta_2''.\label{eq:toymodel2d}
\end{align}
Combining Eqs. \eqref{eq:toymodel1a}--\eqref{eq:toymodel1d} and \eqref{eq:toymodel2a}--\eqref{eq:toymodel2d}
we get
\begin{align}\label{eq:synthesisa}
    J_1=&
    J_1''-\epsilon_0 J_1''(\theta_1''\cos\theta''+\sin\theta'')
        \nonumber \\ &+    
    \frac{\epsilon_0^2}{4}J_1''\left(1
    +2\theta_1''^2
    -2\cos2\theta''
    +2\theta_1''\sin 2\theta''
    \right)
    \nonumber \\&+
    \epsilon_0\epsilon_\omega J_1''[(\theta_1''+\theta'')\cos\theta''-\theta_1''\theta''\sin\theta'']
\nonumber\\
    &+ \mathcal{O}(\epsilon^3),\\
    J_2=& J_2'' + \epsilon_0 J_1'' \theta_1''\cos\theta'' 
            -\frac{\epsilon_0^2}{2}J_1''\left({\theta_1''}^2+\theta_1''\sin2\theta''\right)
        \nonumber \\ &+J_1'' \epsilon_0\epsilon_\omega (\theta''_1\theta''\sin\theta''-\theta_1''\cos\theta'') 
        + \mathcal{O}(\epsilon^3),\\
    \theta_1=&\theta''_1+\epsilon_0\theta''_1\sin\theta'' +
        \frac{\epsilon_0^2}{4} \left(\theta_1''+{\theta_1''}^2\sin2\theta''\right)
    \nonumber \\ &- \epsilon_0\epsilon_{\omega} \theta_1''\theta''\cos\theta    
+  \mathcal{O}(\epsilon^3),\\
    \theta_2=&\theta_2''.
    \label{eq:synthesisd}
\end{align}
Here and in what follows $\mathcal{O}(\epsilon^3)$ denotes $\mathcal{O}(\epsilon^3, \epsilon^2\epsilon_\omega^2, \epsilon\epsilon_{\omega}^2)$. 
The Hamiltonian \eqref{toymodel_hamiltonian} in these canonical variables is
\begin{align}\label{eq:H''}
    \mathcal{H}''(\theta_1'',\theta_2'',J_1'',J_2'') =&
    \left(1 + \epsilon_{\omega 1} -\frac{\epsilon_0^2}4 \right)J_1'' + (1+\epsilon_{\omega 2})J_2'' \nonumber\\&+ \mathcal{O}(\epsilon^3),
\end{align}
implying that the system evolves according to  
\begin{align}\label{eq:toytimestepa}
    J_i'' =& J_{i,0}'' + \mathcal{O}(\epsilon^3)\\
    \label{eq:toytimestepb}
    \theta_i'' =& \theta_{i,0}'' + \omega_i'' t\\
    \omega_i''=&\frac{\partial \mathcal{H}''}{\partial J_i''}=
    \left(
    \begin{array}{c}
        1 + \epsilon_{\omega 1} -\frac{1}4\epsilon_0^2 +\mathcal{O}(\epsilon^3)  \\
        1 + \epsilon_{\omega 2}  
    \end{array}
    \right)
\end{align}
where $i\in\{1,2\}$ and $(\theta_{1,0}'',\theta_{2,0}'',J_{1,0}'',J_{2,0}'')$ are the initial conditions whose values may be obtained from the initial conditions using the inverse transformation 
\begin{align}\label{eq:toyinia}
J_{1,0}'' =& J_{1,0} + \epsilon_0 J_{1,0} (\theta_{1,0}\cos\theta_0
+ \sin\theta_0) \nonumber\\&+ \frac{\epsilon_0^2}{4} J_{1,0} \left(1+2\theta_{1,0}^2-2\theta_{1,0} \sin 2\theta_0\right) + 
\nonumber\\&-\epsilon_0\epsilon_\omega J_1 
[(\theta_1+\theta)\cos\theta - \theta_1\theta\sin\theta ]
+
\mathcal{O}(\epsilon^3),\\
J_{2,0}'' =& J_{2,0} - \epsilon_0 J_1 \theta_{1,0} \cos\theta_0
+\frac{\epsilon_0^2}{2} J_1\left(\theta_{1,0}\sin2\theta_0-\theta_{1,0}^2\right)  
\nonumber\\&+ 
\epsilon_0\epsilon_\omega J_{1,0}\theta_{1,0}(\cos\theta_0 - \theta_0\sin\theta_0)
+\mathcal{O}(\epsilon^3),\\
\theta_{1,0}'' =& \theta_{1,0} - \epsilon_0 \theta_{1,0}\sin\theta_0 + \frac{\epsilon_0^2}{4}\left(\theta_{1,0}-2\theta_{1,0}\cos2\theta_0
 \right.\nonumber\\&+\left. \theta_{1,0}^2\sin2\theta_0\right)+
 \epsilon_{0}\epsilon_{\omega}\theta_{1,0}\theta_0\cos\theta_0+
 \mathcal{O}(\epsilon^3),\\
\theta_{2,0}''=&\theta_{2,0},
\label{eq:toyinid}
\end{align}
where $\theta_0=\theta_{1,0}-\theta_{2,0}$.

\begin{figure*}
\centering
\includegraphics[scale=0.7]{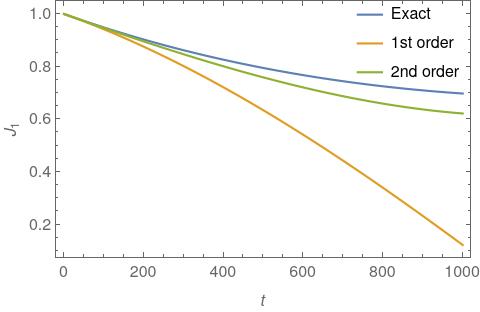}
\includegraphics[scale=0.7]{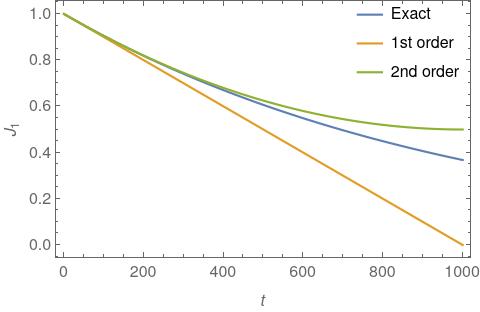}
\caption{The time-evolution of the $J_1$ action for the perturbed harmonic oscillator driven by Eq.~\eqref{toymodel_hamiltonian}. Initial conditions are  $(\theta_{1,0},\theta_{2,0},J_{1,0},J_{2,0})=(1,0,1,1)$ for the left panel and $(0,0,1,1)$ for the right. The perturbing parameters in the left and right panels are $(\epsilon,\epsilon_{\omega 1},\epsilon_{\omega 2})=(0.001,0.002,0.0025)$ and $(0.001,0,0)$, respectively. The orange and green curves show the first-order (Eq.~\ref{eq:toymodel1a}) and second-order (Eq.~\ref{eq:synthesisa}) solutions, respectively. The blue curve shows the exact solution (see Appendix~\ref{app:analytic_toymodel}). Note that the unperturbed orbital frequencies are resonant,  $\omega_{1,0} \approx \omega_{2,0} \approx 1$, and that the errors are small up to $t=\epsilon^{-1} \omega_{1,0}^{-1}$.} 
\label{image}
\end{figure*}

Fig.~\ref{image} shows the time evolution of the $J_1$ momentum for the parameters shown in the figure caption. We compare the exact solution (see Appendix~\ref{app:analytic_toymodel}) with the first- (Eq.~\ref{eq:toymodel1a}) and second-order (Eq.~\ref{eq:synthesisa}) solutions. We note that the case of the right panel is remarkably simple. The exact solution is $J_1=e^{-\epsilon t}$ (see Eq.~\ref{eq:thetaconst}), so the first and second order solutions are $J_1=1-\epsilon t+\frac12 \epsilon^2t^2+\mathcal{O}(\epsilon^3)$. The first three terms of $J_1$'s Taylor series are recovered correctly with the method introduced in this section.

The derivation above may be generalized in a straightforward way to construct higher order methods. However, we note that the $k$th order approximation is valid only if $(\epsilon \omega_{\rm res} t)^{k+1} \lesssim (\epsilon \omega_{\rm res} t)^k$, i.e. $t \lesssim 1/\epsilon$. In order to follow the evolution on longer timescales, the evolution is to be integrated using this method iteratively in steps of $\Delta t\lesssim 1/(\omega_{\rm res}\epsilon)$, where in each step first apply the canonical transformation~\eqref{eq:toyinia}--\eqref{eq:toyinid}, then calculate the time-evolution step~\eqref{eq:toytimestepa}--\eqref{eq:toytimestepb}  for a $\Delta t$ time step, and apply the inverse transformation~\eqref{eq:synthesisa}--\eqref{eq:synthesisd}, and finally set the result to be the initial value of the next iteration step. Since each iteration is generated by integrating the Hamilton's equations of motion exactly for some Hamiltonian, namely that in which the $\mathcal{O}(\epsilon^3)$ terms are neglected, this integrator is symplectic. Symplectic integrators are advantageous as they conserve phase space volume
and all Poincar\'e invariants, and their energy errors typically do not grow systematically with time \citep{BinneyTremaine}. Furthermore the time step may be chosen to be of order $1/\epsilon$, which may be much larger than the inverse frequency. 

\section{Secular dynamics of gravitational triple systems}\label{sec:triplesystems}

We investigate the case of the hierarchical three-body problem. Here, \textit{hierarchical} refers to the small ratio of the separations between the bodies. The potential energy part of the general Hamiltonian for the hierarchical three-body problem is \citep{valtonen}

\begin{align}\label{3bodyhamiltonian}
    \mathcal{H}_\mathrm{pert} =& -\frac{m_1m_2}{r_1} \nonumber\\ &-\frac{m_1m_3}{r_2}\sum_{\ell=0}^{\infty}\left(\frac{-m_2}{m_1+m_2}\right)^\ell \left(\frac{r_1}{r_2}\right)^\ell P_\ell(\cos \psi) \nonumber\\
     &-\frac{m_2m_3}{r_2}\sum_{\ell=0}^{\infty}\left(\frac{m_1}{m_1+m_2}\right)^\ell \left(\frac{r_1}{r_2}\right)^\ell P_\ell(\cos \psi),
\end{align}
where $m_1$ and $m_2$ are the masses of the inner binary, $m_3$ is that of the tertiary, $\mathbf{r}_1$ is the separation vector between the members of the inner binary and $\mathbf{r}_2$ points from the barycenter of the inner binary to the tertiary. $\cos \psi = \mathbf{r_1}\cdot\mathbf{r_2}/(r_1 r_2)$ and $r_1\ll r_2$ due to the hierarchy. 

Now we work out a specific case which is simple enough to illustrate the elimination process introduced in Sec. \ref{sec:general}.
We make the following approximations:
\begin{itemize}
    \item neglect $\ell\geq 3$ multipoles,
    \item $\iota=0$, the mutual inclination vanishes, the objects are in the same plane,
    \item $e_2=0$, outer orbit is circular,
    \item $m_2\ll m_3 \ll m_1$, the central object is massive and the inner binary has a test particle, which implies that the outer binary's semi-major axis, $a_2$, eccentricity, $e_2$, and angular frequency $\omega_2$ are constant, and the mean anomaly follows $l_2=l_{2,0}+\omega_2 t$.
\end{itemize}
With these assumptions, the non-resonant double orbit-averaged Hamiltonian expressed with the Delaunay variables is\footnote{See Eq. (9.30) in \citet{valtonen} with our assumptions.}
\begin{align}\label{eq:valtonen}
    \langle \langle \mathcal{H}_{\rm pert}^{\ell=2}\rangle \rangle =& - \frac{1}{8}m_2m_3\frac{a_1^2}{a_2^3}(2+3e_1^2)
    \nonumber\\
    =& -\frac{1}{8}\frac{m_1m_3^7}{m_2^3}\frac{L_1^4}{L_2^6}\left(5-3\frac{G_1^2}{L_1^2}\right)
    ,
\end{align}
and the secular equations of motion are
\begin{align}\label{sma}
     \dot{L_1}&=-\frac{\partial \langle \langle \mathcal{H}_{\rm pert}^{\ell=2}\rangle \rangle }{\partial l_1}=0\quad\to\quad \dot{a_1} = 0,\\
\label{eccentricity}
     \dot{G_1}&=-\frac{\partial \langle \langle \mathcal{H}_{\rm pert}^{\ell=2}\rangle \rangle }{\partial g_1}=0\quad\to\quad \dot{e_1} = 0,\\
\label{pericenter}
    \dot{g_1} &= \frac{\partial \langle \langle \mathcal{H}_{\rm pert}^{\ell=2}\rangle \rangle }{\partial G_1} = \frac{3}{4}\frac{m_3}{m_1^{1/2}}\frac{a_1^{3/2}}{a_2^3},
\end{align}
where $a_1$ and $a_2$ are the semi-major axes of the inner and outer binaries, respectively, $e_1$ is the inner eccentricity, 
\begin{align}\label{eq:a1a2e1}
    L_1&=m_2\sqrt{m_1a_1}, \nonumber\\
    L_2&=m_3\sqrt{m_1a_2}, \nonumber\\ 
    G_1&=m_2\sqrt{m_1a_1(1-e_1^2)}
\end{align}
are the conjugate canonical momenta to the mean anomalies $l_1$, $l_2$ and the inner argument of pericenter $g_1$. In order to derive these differential equations, one has to average the Hamiltonian over both the inner and outer orbital motions. Even though Eqs. \eqref{sma}-\eqref{pericenter} do not have any apparent divergence within mean motion resonances, they originate from a generating function similar to Eq. \eqref{generating1} that 
diverges suggesting that this may become inaccurate once the inner and outer orbital periods become commensurate. 

Here we demonstrate that the divergence of the generating functions in resonance may be eliminated in this problem using the canonical transformation introduced in Sec.~\ref{sec:general}. For a proof of concept, we present the algorithm through the 1:2 mean motion resonance, although we note that in this case the triple is not hierarchical. First, Eq. \eqref{3bodyhamiltonian} can be rephrased as (see Appendix \ref{appendix} for the derivation) 
\begin{align}\label{eq:fullpert}
\mathcal{H}=&\mathcal{H}_0+\mathcal{H}_1, \\ 
\mathcal{H}_0=&-\frac{m_1^2m_3^3}{2L_2^2} -\frac{m_1^2m_2^3}{2L_1^2},\\ 
\mathcal{H}_1=&-\frac{1}{4}m_2m_3\frac{a_1^2}{a_2^3}  [1+3\cos(2l^*-2l_1)-2e_1\cos l_1  
\nonumber\\ &+  3e_1\cos(2l^*-3l_1)  
- 9e_1\cos(2l^*-l_1) ]
\nonumber\\ &+
   \mathcal{O}\left(e_1^2,e_2,\iota, m_2^2, (a_1/a_2)^4 \right),
\end{align}
where $l^*=l_2-g_1$ and, for the sake of simplicity, we further assume that $e_1 \ll 1$.\footnote{We caution that dropping the terms proportional to $e_1^2$ might cause troubles. For example, if we ignore it in Eq.~\eqref{eq:valtonen}, then the right-hand side of Eq.~\eqref{pericenter} vanishes, hence we miss the pericenter precession. However, as we will show below, the $\sim e_1$ approximation is sufficient to show the effect of the resonance.} At this multipole order, $\ell=2$, we may identify 1:2, 2:3 and 1:1 resonances. However, as we focus on the 1:2 resonance, the other terms can be omitted as long as we are only interested in the terms that systematically grow or decay.
In the 1:2 resonance, the 2:3 and 1:1 terms only induce small periodic oscillations in the orbital elements (however see \citealt{Luo+2016} for the case when $m_3 \sim m_1$). This simplifying assumption does not restrict generality as the non-resonant terms can be accounted for by utilizing the generating function of Eq.~\eqref{eq:oscplusres}.
The truncated perturbing Hamiltonian then consists of a secular and a 1:2 resonant term: 
\begin{align}\label{eq:truncpert}
    \mathcal{H}_{1,\mathrm{tr}}=&\mathcal{H}_{1,\mathrm{s}}+\mathcal{H}_{1,(1:2)}\\
    \mathcal{H}_{1,\mathrm{s}}=& -\frac{1}{4}m_2m_3\frac{a_1^2}{a_2^3}\\
    \mathcal{H}_{1,(1:2)} =& \frac{9}{4}m_2m_3\frac{a_1^2}{a_2^3}e_1\cos(2l^*-l_1).
\end{align}

Note that as the mean anomalies are present only in the $2l^*-l_1=2l_2-2g_1-l_1$ combination, 
\begin{equation}\label{eq:condition}
    \frac{\partial \mathcal{H}_{1,(1:2)}}{\partial l_2} = -2 \frac{\partial \mathcal{H}_{1,(1:2)}}{\partial l_1}.
\end{equation}
The generating functions analogous to Eqs.~\eqref{generating_new1} and \eqref{generating_new2} are
\begin{equation}\label{eq:W12}
    W=-\frac{l_2}{\omega_2}\mathcal{H}_{1,(1:2)}
\end{equation}
and
\begin{equation}\label{eq:W12'}
    W'=\frac{l_2'}{\omega_2}\mathcal{H}'_{1,(1:2)},
\end{equation}
where the prime denotes that the function is expressed with the transformed variables. The transformed variables are generated by $W$ as
\begin{align}\label{eq:L1'}
    L_1'=&\exp(\hat{L}_W)L_1 \approx L_1+[L_1,W] = L_1-\frac{\partial W}{\partial l_1} \nonumber\\ 
    =& L_1 + \frac{9}{4}m_2m_3\frac{a_1^2}{a_2^3} e_1\sin(2l^*-l_1) \frac{l_2}{\omega_2}, \\
    G_1'=&\exp(\hat{L}_W)G_1 \approx G_1+[G_1,W] = G_1-\frac{\partial W}{\partial g_1} \nonumber\\
    =& G_1+\frac{9}{2}m_2m_3\frac{a_1^2}{a_2^3} e_1\sin(2l^*-l_1)\frac{l_2}{\omega_2}, \\
    g_1'=&\exp(\hat{L}_W)g_1\approx g_1+[g_1,W] \approx g_1 + \frac{\partial W}{\partial e_1}\frac{\mathrm{d}e_1}{\mathrm{d}G_1} \nonumber\\
    \approx& g_1+ \frac{9}{4}m_3\frac{a_1^2}{a_2^3} \frac{1}{\sqrt{m_1a_1}}\frac{\cos(2l^*-l_1)}{e_1}\frac{l_2}{\omega_2}.\label{eq:g1'}
\end{align}
The reverse transformation is very similar, but the change with respect to the primed coordinate has an opposite sign (see Eqs~\ref{eq:W12}--\ref{eq:W12'}). When transforming Eq.~\eqref{eq:fullpert}, the variables in the perturbing Hamiltonian can be simply replaced by their primed counterpart at the quadrupole approximation, as they are already multiplied by the small parameter $a_1^2/a_2^2$ (see Eq.~\eqref{eq:F}). The unperturbed Hamiltonian together with the $\mathcal{H}_{1,\rm tr}$ truncated perturbations may be obtained as in Eq.~\eqref{eq:general_hamiltonian}:
\begin{align}
\begin{array}{clc} \mathcal{H}'=&-\dfrac{m_1^2m_2^3}{2L_1'^2}-\dfrac{m_1^2m_3^3}{2L_2'^2}&[\mathrm{I}]\nonumber\\[2ex]    &+\dfrac{m_1^2m_2^3}{L_1'^3}\left(-\dfrac{l_2'}{\omega_2}\dfrac{\partial \mathcal{H}'_{1,(1:2)}}{\partial l_1'}\right) &[\mathrm{II}]\nonumber\\[2ex] &+\dfrac{m_1^2m_3^3}{L_2'^3}\left(-\dfrac{l_2'}{\omega_2}\dfrac{\partial \mathcal{H}'_{1,(1:2)}}{\partial l_2'}\right)&[\mathrm{III}]\nonumber\\[2ex]
    &+\dfrac{m_1^2m_3^3}{L_2'^3}\left(-\dfrac{1}{\omega_2}\mathcal{H}'_{1,(1:2)}\right)&[\mathrm{IV}]\nonumber\\[2ex]
    &+\mathcal{H}'_{1,(1:2)}&[\mathrm{V}]\\[2ex]
    &+\mathcal{H}'_\mathrm{s},
\end{array}
\end{align}
where in the first-order Taylor expansion we used $L_k \approx L_k'+[L_k',W']=L_k'-\partial W'/\partial l_k'$. Using the definition of orbital frequencies
\begin{equation}
    \omega_1=\frac{\partial \mathcal{H}_0}{\partial L_1}=\frac{m_1^2m_2^3}{L_1'^3}+\mathcal{O}\left(\frac{a_1^2}{a_2^2}\right)
\end{equation}
and
\begin{equation}
    \omega_2=\frac{\partial \mathcal{H}_0}{\partial L_2}=\frac{m_1^2m_3^3}{L_2'^3}+\mathcal{O}\left(\frac{a_1^2}{a_2^2}\right),
\end{equation}
and the fact that from Eq.~\eqref{eq:condition} $\frac{\partial}{\partial l_2'}=-2\frac{\partial}{\partial l_1'}$, rows IV and V mutually cancel and the Hamiltonian simplifies as
\begin{align}
    \mathcal{H}'=&-\frac{m_1^2m_2^3}{2L_1'^2}-\frac{m_1^2m_3^3}{2L_2'^2}\nonumber\\
    &-\frac{l_2'}{\omega_2}\frac{\partial \mathcal{H}'_{1,(1:2)}}{\partial l_2'}\left(\omega_1-2\omega_2\right)
    +\mathcal{H}'_\mathrm{s}\nonumber\\
    =& -\frac{m_1^2m_2^3}{2L_1'^2}-\frac{m_1^2m_3^3}{2L_2'^2}+\mathcal{H}'_\mathrm{s},
\end{align}
where the term in the parenthesis vanishes because of the 1:2 resonant condition. As it is expected, $\mathcal{H}'_\mathrm{s}=\mathcal{H}_\mathrm{pert}^{\ell=2}$ if $e_1^2\approx 0$. The new equations of motions in the transformed variables are:
\begin{align}
    \dot{L_1'}&=0, \\
    \dot{G_1'}&=0, \\
    \dot{g_1'}&=0.
\end{align}
These equations are orbit-averaged in the sense that the non-resonant oscillating terms have been eliminated from the Hamiltonian. Integrating them gives
\begin{align}\label{sma_new}
    L_1'&=\mathrm{const.}, \\
\label{eccentricity_new}
    G_1'&=\mathrm{const.}, \\
\label{pericenter_new}
    g_1' &= \mathrm{const}.
\end{align}
These equations are nearly identical to Eqs. \eqref{sma}-\eqref{pericenter}, the only difference is the absence of pericenter precession, which is the result of ignoring the $\sim e_1^2$ terms for simplicity. We stress, however, that even though these equations are formally the same, they are derived by a completely different generating function, which avoid divergences in mean motion resonances, and these variables are related to the $(a,e,g)$ orbital elements differently. 

Eqs. \eqref{sma_new}-\eqref{pericenter_new} may be expressed with the $(a,e,g)$ orbital elements using Eqs.~\eqref{eq:a1a2e1} and \eqref{eq:L1'}--\eqref{eq:g1'} as 
\begin{align}
     a_1 &= \mathrm{const.}-\frac{9}{2}\frac{m_3a_1^{5/2}}{m_1^{1/2}a_2^{3}}e_1\sin(2l^*-l_1)\frac{l_2}{\omega_2}, \\
    e_1 &= \mathrm{const.}+\frac{9}{4}\frac{m_3a_1^{3/2}}{m_1^{1/2}a_2^3}\sin(2l^*-l_1)\frac{l_2}{\omega_2}, \\
    g_1 &= \mathrm{const.}-\frac{9}{4}\frac{m_3a_1^{3/2}}{m_1^{1/2}a_2^3} \frac{\cos(2l^*-l_1)}{e_1} \frac{l_2}{\omega_2}.
\end{align}
Substituting $l_2=l_{2,0}+\omega_2 t$ and setting the $\mathrm{const.}$ terms properly, it yields that

\begin{align}
     a_1 &= a_{1,0}-\frac{9}{2}\frac{m_3a_1^{5/2}}{m_1^{1/2}a_2^{3}}e_1\sin(2l^*-l_1)\,t, \\
    e_1 &= e_{1,0}+\frac{9}{4}\frac{m_3a_1^{3/2}}{m_1^{1/2}a_2^3}\sin(2l^*-l_1)\,t, \\
    g_1 &= g_{1,0}-\frac{9}{4}\frac{m_3a_1^{3/2}}{m_1^{1/2}a_2^3} \frac{\cos(2l^*-l_1)}{e_1} \,t,
\end{align}
where the time dependence is also implicit in the variables $(l^*,l_1,e_1,a_1)$.

\begin{figure*}
\centering
\includegraphics[scale=0.35]{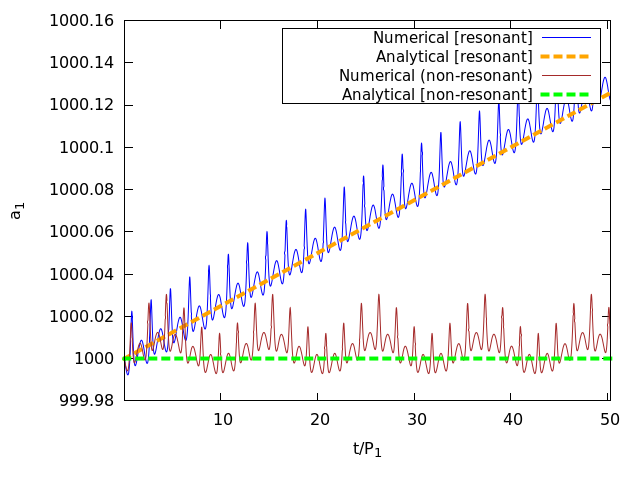}
\includegraphics[scale=0.35]{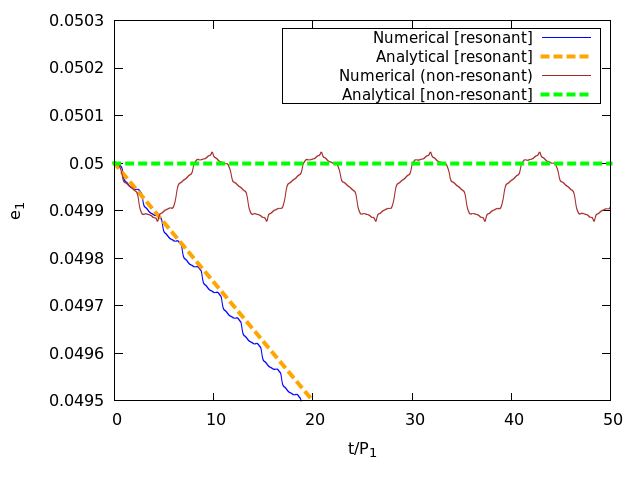}
\includegraphics[scale=0.35]{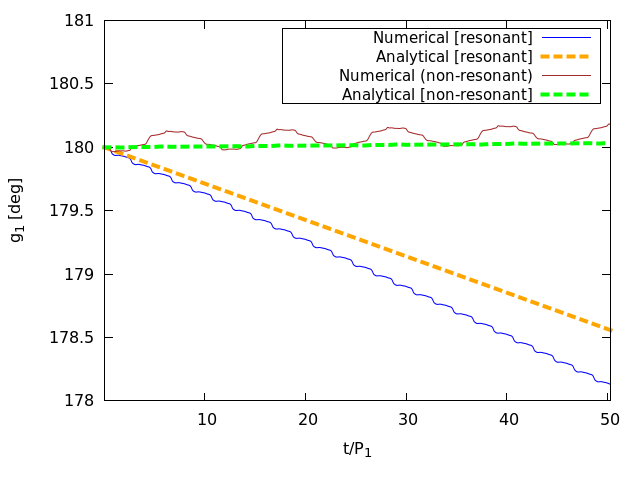}
\caption{Comparison of the analytical and numerical results within 1:2 mean motion resonance (blue and orange curves) and slightly out of it ($P_1/P_2=1/2.2$; brown and green curves). The masses are $m_1/m_2=10^6$, $m_3/m_2=10$, the semi-major axes are $a_1=1000$ and $a_2=1587.4$ (arbitrary units), the inner eccentricity is $e_1=0.05$, the initial mean anomalies are $l_{1,0}=45^\circ$, $l_{2,0}=0$. While the non-resonant analytical solution clearly misses the systematic secular change, the analytical resonant solution is a better approximation. The discrepancy is mostly due to neglecting the octupole and higher multipoles.}
\label{g1}
\end{figure*}

In Figure~\ref{g1} we plot the evolution of the orbital elements both in the resonant ($P_1/P_2=1/2$) and in the non-resonant ($P_1/P_2=1/2.2$) case. The blue and brown curves are the numerical result, while the thick dashed green and orange curves are the fits from the non-resonant and resonant equations, respectively. We observe qualitative agreement between the direct numerical and the analytical results using the resonant generating function. The discrepancy between the analytical and numerical results is unsurprising. It originates from the approximations we used, especially from the quadrupole assumption. The relative error may be estimated through the ratio of the octupole and the quadrupole terms which in the 1:2 resonance is
\begin{equation}
    \frac{(a_1/a_2)^3}{(a_1/a_2)^2}=\frac{a_1}{a_2} = \left(\frac{1}{2}\right)^{2/3} \approx 0.63. 
\end{equation}
To follow the evolution of the system more accurately on longer timescales, the multipole expansion must be carried out to higher orders,  and one must use a smaller time step with an iterative symplectic integration scheme described in Sec. \ref{sec:toymodel}. We leave this to future work.

\section{Discussion}\label{sec:discussion}
In this paper we proposed a novel canonical transformation to eliminate the quick angle variables in dynamical systems that admit action-angle variables to leading order and which are subject to resonant perturbations. For a proof of concept, we applied this technique to coupled resonant harmonic oscillators and to the gravitational restricted three-body problem on nearly circular, coplanar orbits in mean motion resonance.

This transformation defines a new set of canonical variables explicitly as functions of the orbital elements, in which the perturbed Hamilton's equations of motion may be integrated trivially. The transformed variables evolve according to the unperturbed equations of motion.
Specifically, we have shown that if the perturbations in the original Hamiltonian were of order $\epsilon$, the non-integrable part of the perturbation in the transformed variables becomes $\propto\epsilon^2$. We have shown that repeated applications of similar canonical transformations may be used to extend the integrable part to arbitrary accuracy in a series of powers of $\epsilon$. Since Hamilton's equations may be integrated in the transformed variables with a large time step $\Delta t\propto 1/\epsilon$. The iterative application of the canonical transformation, time evolution, and reverse transformation yields an efficient symplectic integrator to simulate the time evolution of the system.

We note that this algorithm cannot be directly applied to systems of 3 (uneven) degrees of freedom (for example, to Laplace resonances), since resonant terms in the Hamiltonian cancel each other in pairs (see rows III and IV in Eq. \eqref{eq:general_hamiltonian}). 

However, for even degrees of freedom, the applicability of this algorithm is not restricted to the simplifying assumptions adopted in the toy models presented here. In the future we plan to further develop this method by (i) relaxing the constraints of the orbital elements to make it applicable for general non-zero inclinations, arbitrary inner and outer eccentricities,  (ii) incorporating terms of higher orders in the Hamiltonian, both in $(a_1/a_2)$ and $e$; (iii) including general relativistic effects, apsidal precession and gravitational wave radiation, and explore Kozai-Lidov oscillations for triple systems that sweep through resonances.

\acknowledgments{We thank Smadar Naoz, Bálint Érdi, John Magorrian and Scott Tremaine for helpful comments. We are also grateful to Mária Kolozsv\'ari for help with logistics and administration related to the research. This work received funding from the European Research Council (ERC) under the European Union's Horizon 2020 research and innovation program under grant agreement No 638435 (GalNUC). Barnabás Deme was supported by the \'UNKP-20-3-II New National Excellence Program of the Ministry for Innovation and Technology from the source of the National Research, Development and Innovation Fund.}

\appendix 
\section{Derivation of the resonant Hamiltonian} \label{appendix}

Here we derive the 1:2 resonant Hamiltonian for the gravitational three-body problem up to quadrupole accuarcy. 

The cosine of the angle between $\mathbf{r}_1$ and $\mathbf{r}_2$ is 

\begin{align}\label{cospsi}
\cos \psi &= \frac{\mathbf{r}_1\cdot \mathbf{r}_2}{r_1r_2} = \cos\left(l_2-(g_1 + v_1)\right) = \cos \left( (l_2-g_1) - v_1 \right) \nonumber\\ &=\cos (l^*-v_1)= \cos l^* \cos v_1 + \sin l^* \sin v_1,
\end{align}
where $v_1$ is the true anomaly of the inner binary and the other notations are the same as before. The argument of the inner periapsis is present only in the combination $l^* = l_2-g_1$, so $g_1$ disappears when we average over $l^*$ (or $l_2$). This implies that the inner eccentricity ($e_1$) remains constant on secular timescale. Now we have to express the true anomaly by the mean one, for which we use that $e \ll 1$ and $e^2 \approx 0$. In the $\mathcal{O}(e)$ approximation the Kepler equation is modified as
\begin{equation}
    E_1=l_1+e_1 \sin l_1,
\end{equation}
where $E_1$ is the eccentric anomaly and from which it follows that
\begin{equation}
    \sin E_1 = \sin (l_1 + e_1 \sin l_1 ) = \sin l_1 + e_1 \sin l_1 \cos l_1,
\end{equation}
\begin{equation}\label{cosE}
    \cos E_1 = \cos (l_1 + e_1 \sin l_1) = \cos l_1 - e_1 \sin^2 l_1.
\end{equation}
Converting the eccentric anomaly to the true one, we obtain 
\begin{align}\label{sinv}
    \sin v_1 =& \frac{\sqrt{1-e_1^2}\sin E_1}{1-e_1\cos E_1}  = \sin E_1 (1+e_1 \cos l_1) +\mathcal{O}(e_1^2)= \sin l_1 + e_1\sin(2l_1)+\mathcal{O}(e_1^2).\\
\label{cosv}
    \cos v_1 =& \frac{\cos E_1 -e_1 }{1-e_1\cos E_1} = (\cos E_1 -e_1)(1+e_1\cos l_1)+\mathcal{O}(e_1^2) = -e_1 + \cos l_1 + e_1 \cos (2l_1) \HH+\mathcal{O}(e_1^2).
\end{align}
Substituting Eqs. \eqref{sinv} and \eqref{cosv} into \eqref{cospsi} we get
\begin{align}
\cos \psi =& \cos l^* \left[\cos l_1 -e_1 + e_1 \cos (2l_1)\right] 
+ \sin l^* \left[ \sin l_1 + e_1 \sin (2l_1) \right]
+\mathcal{O}(e_1^2)
\nonumber\\
=& 
\cos (l^*-l_1) -e_1 \cos l^*  +  e_1 \cos (l^*-2l_1)
+\mathcal{O}(e_1^2)
\end{align}
and from Eq. \eqref{cosE} 
\begin{equation}
    r_1=a_1(1-e_1\cos l_1)+\mathcal{O}(e_1^2).
\end{equation}
As the outer orbit is circular, $r_2=a_2$. With all this preparation, the perturbing Hamiltonian in Eq. \eqref{3bodyhamiltonian} reads (up to quadrupole order and with $m_2 \ll m_1$)
\begin{align}
    \mathcal{H}_1=&-\frac{1}{2}\frac{m_1m_2m_3}{m_1+m_2}\frac{r_1^2}{a_2^3}\left( 3\cos^2 \psi -1 \right)  \nonumber\\
    =& -\frac{1}{4}m_2m_3\frac{a_1^2}{a_2^3} \left[ 1+3\cos(2l^*-2l_1) + 3e_1\cos(2l^*-3l_1) -2e_1\cos l_1 -9e_1\cos(2l^*-l_1) \right]
    +\mathcal{O}\left(e_1^2,e_2,\iota, m_2^2 \right).
\end{align}

The term that corresponds to the 1:2 mean motion resonance is
\begin{equation}
    \mathcal{H}_{1,(1:2)} = \frac{9}{4}m_2m_3\frac{a_1^2}{a_2^3}e_1\cos(2l^*-l_1).
\end{equation}

\section{Exact solution to the perturbed harmonic oscillator}
\label{app:analytic_toymodel}
Here we present the exact solution for the Hamiltonian Eq.~\eqref{toymodel_hamiltonian} \begin{equation}\label{toymodel_hamiltonianapp}
    \mathcal{H}=(1+\epsilon_{\omega,1})J_1+
    (1+\epsilon_{\omega,2})J_2 +\epsilon J_1 \sin(\theta_1-\theta_2),
\end{equation}
Introduce the canonical transformation
\begin{align}\label{eq:appcanonical}
\left(
\begin{array}{c}
     \varphi  \\
     \theta  \\
     U  \\
     V
\end{array}
\right)
=
\left(
\begin{array}{cccc}
    \frac12 & \frac12   & 0 & 0 \\
     1      & -1        & 0 & 0 \\
     0      & 0         & 1 & 1  \\
    0       & 0         &\frac12 & -\frac12
\end{array}
\right)
\left(
\begin{array}{c}
     \theta_1  \\
     \theta_2  \\
     J_1  \\
     J_2
\end{array}
\right)\,,
\quad 
\left(
\begin{array}{c}
     \theta_1  \\
     \theta_2  \\
     J_1  \\
     J_2
\end{array}
\right)
=
\left(
\begin{array}{cccc}
     1  & \frac12   & 0 & 0 \\
     1  & -\frac12  & 0 & 0 \\
     0  & 0         & \frac12 & 1  \\
     0  & 0         &\frac12 & -1
\end{array}
\right)
\left(
\begin{array}{c}
     \varphi  \\
     \theta  \\
     U  \\
     V
\end{array}
\right)
\end{align}
so that 
\begin{equation}
    \mathcal{H} = c U + \epsilon_{\omega} V + \epsilon \left(\frac{U}{2}+V\right) \sin \theta
\end{equation}
where $c = 1+\frac12(\epsilon_{\omega 1}+\epsilon_{\omega 2})$ and $\epsilon_{\omega} = \epsilon_{\omega 1}-\epsilon_{\omega 2}$. Hamilton's equations of motion are
\begin{align}
    \frac{d \varphi}{dt} &= \frac{\partial \mathcal{H}}{\partial U} = c + \frac{\epsilon}{2}\sin \theta\\
    \frac{d \theta}{dt} &= \frac{\partial \mathcal{H}}{\partial V} = \epsilon_{\omega} + \epsilon \sin \theta\\
    \frac{d U}{dt} &= -\frac{\partial \mathcal{H}}{\partial \varphi} = 0\\
    \frac{d V}{dt} &= -\frac{\partial \mathcal{H}}{\partial \theta} = -\epsilon\left(\frac{U}{2}+V\right)\cos \theta\label{eq:dV/dt}
\end{align}
This shows that $U$ is a constant, and we may integrate the equation for $\theta$ and express all other phase space variables with $\theta$. A trivial solution is the case when $\epsilon\sin\theta_0=-\epsilon_{\omega}$, then $\theta=\theta_0$ for all times and the equations of motions may be integrated to give
\begin{align} 
\left(
\begin{array}{c}
     \varphi  \\
     \theta  \\
     U  \\
     V
\end{array}
\right)
=
\left(
\begin{array}{c}
    \varphi_0 + c t  \\
     0     \\
     U_0      \\
    \left(\frac{U_0}2 + V_0\right)
    e^{-\epsilon t \cos \theta_0}
    -\frac{U_0}2      
\end{array}
\right)\,,
\quad
\left(
\begin{array}{c}
     \theta_1  \\
     \theta_2  \\
     J_1  \\
     J_2
\end{array}
\right)
=
\left(
\begin{array}{c}
     \theta_{2,0} + c t  \\
     \theta_{1,0} + c t  \\
     J_{1,0} e^{-\epsilon t \cos \theta_0}  \\
     J_{2,0} + J_{1,0}(1-e^{-\epsilon t \cos \theta_0})
\end{array}
\right)
\mathrm{~~if~~} \epsilon\sin\theta_0=-\epsilon_{\omega}\,.
\label{eq:thetaconst}
\end{align}
Otherwise we will assume that $\epsilon\sin\theta_0\neq -\epsilon_{\omega}$. Then
\begin{align}\label{eq:t(theta)}
    t(\theta) &= \int_{\theta_0}^{\theta} \frac{d\theta}{d\theta/dt}  = \int_{\theta_0}^{\theta} \frac{d \theta'}{\epsilon_{\omega} + \epsilon\sin \theta'} = 
    \left\{
    \begin{array}{cc}
             n t_p+\left.\dfrac{2\,\mathrm{sign}(\epsilon_{\omega})}{\sqrt{\epsilon_{\omega}^2 - \epsilon^2}} \tan^{-1}\left(\dfrac{\sin\left(\dfrac{\theta'+\alpha}{2}\right)}{\cos\left(\dfrac{\theta'-\alpha}{2}\right)}\right)\right|_{\theta_0}^{\theta} & \mathrm{~if~}|\epsilon|\leq|\epsilon_{\omega}|
  \\[2ex]
\left.\dfrac{\mathrm{sign}(\epsilon)}{\sqrt{\epsilon^2 - \epsilon_{\omega}^2}}
\ln\left(\dfrac{\sin\left(\dfrac{\theta'+\alpha}{2}\right)}{\cos\left(\dfrac{\theta'-\alpha}{2}\right)}\right)\right|_{\theta_0}^{\theta}         &  \mathrm{~if~}|\epsilon_{\omega}|\leq |\epsilon|
    \end{array}
    \right.
\end{align}
where $t_p=t(2\pi)-t(0)$, $n$ is an integer, and $\alpha = \sin^{-1}(\epsilon/\epsilon_{\omega})$ if $|\epsilon|\leq|\epsilon_{\omega}|$ and $\alpha = \sin^{-1}(\epsilon_{\omega}/\epsilon)$ otherwise if $|\epsilon_{\omega}|\leq|\epsilon|$. Note that in both cases this may be inverted analytically to give $\theta(t)$ in a closed form, but the resulting expression is complicated and we do not show it here. If $|\epsilon|\leq|\epsilon_{\omega}|$ then, depending on the sign of $\epsilon_{\omega}$, $\theta$ either increases or decreases monotonically for all times without bounds\footnote{However note that $\theta\equiv \theta+2n\pi$ if $n$ is an integer}, otherwise if $|\epsilon|\geq|\epsilon_{\omega}|$ then, depending on the sign of $\epsilon\sin\theta_0$, $\theta$ increases or decreases monotonically such that $\sin \theta$ approaches $-\epsilon_{\omega}/\epsilon$ as $t\rightarrow\infty$.
Now to solve for the evolution  of $V$, divide $dV/dt$ by $d\theta/dt$
\begin{align}
    \frac{dV/dt}{d\theta/dt} &=     \frac{dV}{d\theta} =  -\frac{\epsilon\left(\frac{U}{2}+V\right)\cos \theta}{\epsilon_{\omega} + \epsilon \sin \theta}\,.
\end{align}
This is a separable differential equation since $U$ is a constant.
\begin{align}
    \int_{V_0}^{V} \frac{dV'}{\frac12U_0+V'} = \ln\left(\frac{\frac12U_0 + V}{\frac12U_0 + V_0}\right) &=  -\int_{\theta_0}^{\theta} d\theta' \frac{\epsilon\cos \theta'}{ \epsilon_{\omega} + \epsilon \sin \theta'} =  - \ln \left(\frac{ \epsilon_{\omega} + \epsilon\sin\theta}{ \epsilon_{\omega} + \epsilon\sin\theta_0}\right) \\
    V(\theta) &=
    \left(\frac{U_0}2 + V_0\right)
    \frac{ \epsilon_{\omega} + \epsilon\sin\theta_0}{ \epsilon_{\omega} + \epsilon\sin\theta}-\frac{U_0}2\label{eq:V}\,,
\end{align}
Finally to find the evolution of $\varphi$,  divide $d\varphi/dt$ by $d\theta/dt$
\begin{equation}
      \frac{d\varphi/dt}{d\theta/dt} =        \frac{d\varphi}{d\theta} = 
\frac{c+\frac{\epsilon}2\sin \theta}{\epsilon_{\omega} + \epsilon \sin \theta} = \frac12  + 
      \frac{c - \frac{\epsilon_{\omega}}{2}}{\epsilon_{\omega} + \epsilon \sin \theta} 
      = \frac12  + 
      \left(c - \frac{\epsilon_{\omega}}{2}\right)\frac{dt}{d\theta}
\end{equation}
This may be integrated with respect to $\theta$ to give
\begin{equation}
    \varphi = \varphi_0 + \frac{\theta-\theta_0}{2}
    + \left(c - \frac{\epsilon_{\omega}}{2}\right)t
\end{equation}
Substituting into Eq.~\eqref{eq:appcanonical} gives the parameteric solution in the original variables:
\begin{equation}
\left(
\begin{array}{c}
     \theta_1  \\
     \theta_2  \\
     J_1  \\
     J_2
\end{array}
\right)
=
\left(
\begin{array}{c}
     \theta_{1,0} +\left(1+ \epsilon_{\omega 2}\right)t +\theta - \theta_0   \\
     \theta_{2,0}+\left(1+ \epsilon_{\omega 2}\right)t  \\
     J_{1,0}\dfrac{ \epsilon_{\omega} + \epsilon\sin\theta_0}{ \epsilon_{\omega} + \epsilon\sin\theta}  \\[2ex]
      J_{2,0} + J_{1,0}\dfrac{ \epsilon(\sin\theta-\sin\theta_0)}{ \epsilon_{\omega} + \epsilon\sin\theta}
\end{array}
\right)\,,\mathrm{~~if~~} \epsilon\sin\theta_0\neq -\epsilon_{\omega}\,.    
\end{equation}
where $t(\theta)$ is given by Eq.~\eqref{eq:t(theta)}. Note that for $|\epsilon_{\omega}| > |\epsilon|$ $\sin\theta$ is oscillatory in the full range $-1$ and $1$, and for $|\epsilon_{\omega}| \leq |\epsilon|$ it changes monotonically and so $J_1$ and $J_2$ exhibit bounded oscillations and for $|\epsilon_{\omega}| \leq |\epsilon|$, the denominator asymptotically vanishes so $J_1/J_{1,0}\rightarrow \infty$ and $J_2/J_{1,0}\rightarrow -\infty$. For $\epsilon\sin\theta_0=-\epsilon_{\omega}$ the evolution is given by Eq.~\eqref{eq:thetaconst}.

\bibliography{cit}{}
\bibliographystyle{aasjournal}

\end{document}